\documentstyle[emulateapj,psfig,flushrt,tighten]{article}

\def\etal{{et al.\ }}
\def\gsim{ \lower .75ex \hbox{$\sim$} \llap{\raise .27ex \hbox{$>$}} }
\def\lsim{ \lower .75ex\hbox{$\sim$} \llap{\raise .27ex \hbox{$<$}} }
\def\msun{\,{\rm M_\odot}}

\righthead{Gravitational Waves from Black Hole Binaries}
\lefthead{Sesana \etal}
\slugcomment{ApJ, in press}
\makeatletter

\newenvironment{figurehere}
  {\def\@captype{figure}}
  {}
\makeatother

\begin{document}

\title{The gravitational wave signal from massive black hole binaries and 
its contribution to the {\it LISA} data stream}

\author{Alberto Sesana\altaffilmark{1}, Francesco Haardt\altaffilmark{1},
Piero Madau\altaffilmark{2}, \& Marta Volonteri\altaffilmark{2}}

\altaffiltext{1}{Dipartimento di Fisica \& Matematica, Universit\'a dell'Insubria, via
Valleggio  11, 22100 Como, Italy.}
\altaffiltext{2}{Department of Astronomy \& Astrophysics, University of
California, 1156 High Street, Santa Cruz, CA 95064.}

\begin{abstract}
Massive black hole binaries, with masses in the range
$10^{3}-10^{8} M_\odot$, are expected to be the most powerful sources of gravitational 
radiation at mHz frequencies, and hence are among the primary targets for the
planned {\it Laser Interferometer Space Antenna} ({\it LISA}). We extend
and refine our previous analysis (Sesana \etal 2004), detailing
the gravitational wave signal expected from a cosmological
population of massive black hole binaries. As done in our previous
paper, we follow the merger history of dark matter halos, the dynamics 
of the massive black holes  they host, and their growth via gas
accretion and binary coalescences in a $\Lambda$CDM cosmology.
Stellar dynamical processes dominates the orbital evolution of black
hole binaries at large separations, while gravitational wave emission 
takes over at small radii, causing the final coalescence of the pairs.
We show that the GW signal from this population, in a 3 year {\it LISA}
observation, will be resolved into $\simeq 90$ discrete events
with $S/N \geq 5$, among which $\simeq 35$ will be observed above
threshold until coalescence. These ``merging events" involve
relatively massive binaries, $M\sim 10^5 \msun$, in the redshift
range $2\lsim z \lsim 6$. The remaining $\simeq 55$ events come
from higher redshift, less massive binaries ($M \sim 5\times 10^3 \msun$
at $z \gsim 6$) and, although their $S/N$ integrated over the duration of
the observation can be substantial, the final coalescence phase is at too
high frequency to be directly observable by space--based interferometers such as 
{\it LISA}. {\it LISA} will 
be able to detect a fraction $\gsim 90$\% of all the coalescences of massive 
black hole binaries occurring at $z\lsim 5$.   
The residual confusion noise from unresolved 
massive black hole binaries is expected to be at least an
order of magnitude below the estimated stochastic noise.

\end{abstract}


\section{Introduction}

Gravitational radiation, 
described as a tensor perturbation to the metric travelling at the speed of light, 
is a natural consequence of Einstein's general relativity. 
It has been recognised (e.g., Thorne 1987) that black hole binaries are among the most important sources 
of gravitational waves (GW), both for ground based interferometers such as {\it LIGO} 
(Abramovici \etal 1992) and {\it VIRGO} (Bradaschia \etal 1990), and 
for the planned {\it Laser Interferometer Space Antenna} ({\it LISA}, Bender \etal 1994). 

Interferometers operate as all-sky monitors, and the data streams collect 
the contributions from a large number of sources belonging to 
different cosmic populations. 
A precise determination of stochastic GW backgrounds from different 
classes of astrophysical objects is therefore crucial to interpret the 
data. While a GW background may provide information on the number density, 
redshift evolution, and mass function of the emitting population,    
confusion noises add to the instrumental noise limiting 
the possibility of detecting other class of objects. Moreover, to optimize 
the subtraction of resolved sources from the data stream, it is important to 
have a detailed description of the expected rate, duration, amplitude, 
and waveforms of events.  
    
{\it LISA} will operate in the frequency range 0.01 mHz - 1 Hz, where GW 
emission from a cosmological population of massive black hole binaries 
(MBHBs) is expected to be important (Haehnelt 1994). Today, massive black 
holes (MBHs) are ubiquitous in the
nuclei of nearby galaxies (see, e.g., Magorrian et al. 1998). If MBHs were
also common in the past, and if their host
galaxies experience multiple mergers during their lifetime, as dictated by
popular cold dark matter hierarchical cosmologies, then MBHBs 
will inevitably form in large numbers during cosmic history. The 
formation and evolution of MBHs has been investigated recently by several 
groups (e.g., Menou, Haiman \& Narayanan 2001; Volonteri \etal 2003), 
and the expected GW signal from inspiraling MBH binaries has been first 
discussed by Rajagopal \& Romani (1995), and recently by Jaffe \& Backer (2003), Wyithe \& Loeb (2003), 
Sesana \etal (2004, hereafter Paper I), and Enoki \etal (2004). 

In Paper I we computed the GW background from MBHBs and the number 
of coalescences  observable by {\it LISA} in a 3-year mission,  
adopting the scenario for the assembly and growth of MBHs proposed
by Volonteri \etal (2003a,b). 
In such model, ``seed'' holes are placed within rare
high-density regions (minihalos) above the cosmological Jeans and cooling
mass at redshift 20. Their evolution is followed through Monte
Carlo realizations of the halo merger hierarchy combined with semi-analytical
descriptions of the main dynamical processes, such as dynamical friction
against the dark matter background, the shrinking of MBH binaries via
three-body interactions, their coalescence driven by the emission of
gravitational waves, and the recoil associated with the non-zero
net linear momentum carried away by GWs in the coalescence of two unequal mass
black holes (the ``gravitational rocket''). Major halo mergers lead to
MBH fueling and trigger quasar activity. In this paper we use the same model
to provide a more detailed characterization of the GW signal from 
inspiraling MBHBs. Their contribution to the {\it LISA} data stream is 
twofold: unresolved sources will give origin to confusion noise to be 
compared to instrumental noise and other astrophysical stochastic 
backgrounds (e.g. from white dwarf binaries, Farmer \& Phinney 2003), while
resolved inspiraling binaries will probe gravity in extreme conditions 
(e.g., Vecchio 2004). Confusion noise and resolved sources should provide 
different cosmological information. The former, produced by a large 
number of unresolved MBHBs, will trace light MBHBs at very high redshift, 
placing constraints on black hole formation scenarios prior to 
the reionization epoch; the latter will be a formidable tool to follow 
the cosmic evolution of MBHs and the formation and dynamics of MBH binaries 
following galaxy mergers.  

The plan is as follows. In \S~2 we review the basics of the detection of 
GW from MBHBs, defining observable quantities such as the characteristic 
strain amplitude, signal-to-noise ratio, and source detection 
rate. In \S~3 we briefly summarize our scenario for the cosmological
evolution of galaxy halos and associated holes. In \S~4 we present confusion 
noise levels and source number counts. Finally, in \S 5 we discuss our 
results.    

\section{Gravitational wave signals}

\subsection{Bursts and periodic events}

Following Thorne (1996), an interferometer can be characterized by
two different sensitivity curves, depending on the type of signal
one expects to detect, i.e. a ``burst" or a ``periodic" GW source. A
burst, a short-lived signal whose waveform can be utterly
complicated, can be described in terms of a characteristic strain
amplitude $h_c$ at the observed frequency $f_c \sim 1/\Delta
t_s$, where $\Delta t_s$ is the duration of the signal (Thorne
1987). The spread of the power spectrum around $f_c$ will be
$\Delta f \sim f_c$. At the other extreme, a perfectly periodic
source emits, for the entire duration of the observation, at a
fixed frequency $f$. The power spectrum will be peaked at $f$,
with a spread $\Delta f \simeq f/N$, where $N$ is the number of
wave cycles clipped into the observation. In this respect, a burst
can be thought as a single complete waveform with $f=f_c$. In the
case of a periodic signal, the interferometer sensitivity is
increased by the fact that, across the observing interval $\tau$,
the signal is repeated $f\tau$ times.

The sensitivity to bursts ($h_B$) and to periodic signals ($h_P$) are related by:
\begin{equation}
h_P(f)=\frac{h_B(f)}{\sqrt{f\tau}}.
\label{eqburstperiodic}
\end{equation}
In Figure \ref{sensitivity} the two curves $h_B$ and $h_P$ are
compared for an assumed 3-year {\it LISA} observation. The curves are
obtained combining the {\it LISA} single-arm Michelson sensitivity curve
(taken from the URL
www.srl.caltech.edu/$\sim$shane/sensitivity) with the recent
analysis of the {\it LISA} instrumental noise below $10^{-4}$ Hz (Bender
2003, extended from $3\times 10^{-6}$ Hz to $1\times 10^{-6}$ Hz
with a constant slope).

Consider now a periodic signal of finite duration, with
strain amplitude $h$. The total energy carried by the wave will be
proportional to the number of wave cycles $n$ spent at that
particular frequency. The quantity to be compared with $h_B$ is
then the ``characteristic" strain $h_c \equiv h\sqrt{n}$.
\begin{figurehere}
\vspace{0.5cm}
\centerline{\psfig{figure=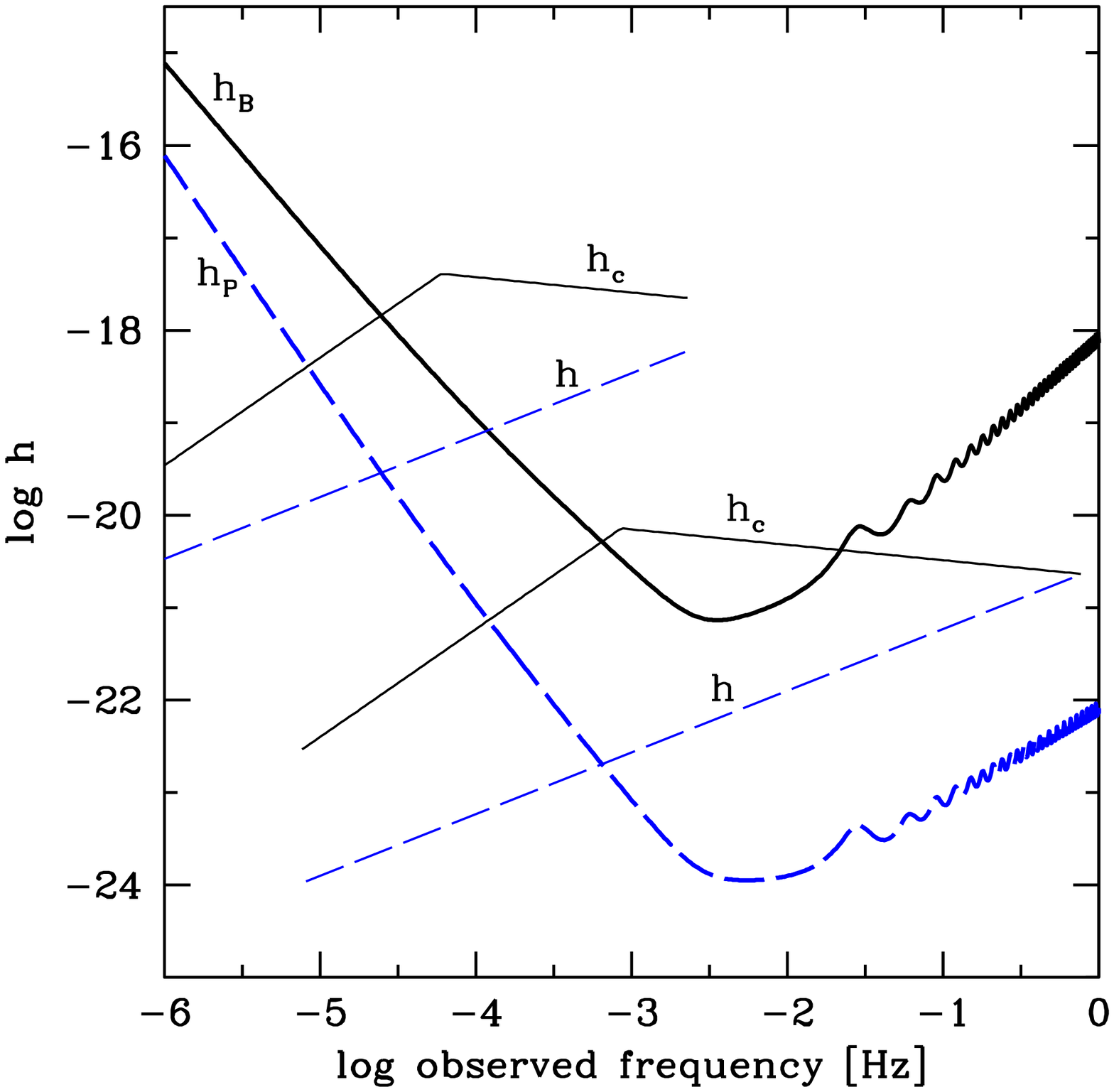,width=3.6in}}
\vspace{-0.0cm}
\caption{\footnotesize 
{\it LISA} single-arm Michelson sensitivity curve
to bursts ({\it thick solid line}) and periodic signals ({\it thick dashed line})
in a 3-year mission.
Data are obtained from www.srl.clatech.edu/$\sim$shane/sensitivity, and Bender (2003).
The strain $h$ (dashed lines) and characteristic strain $h_c$ (solid lines) for 
a MBHB with $M_2=0.1M_1=10^5 M_{\odot}$ at $z=1$ (upper lines), and $M_2=M_1=10^3 M_{\odot}$ 
at $z=7$ (lower lines) are also shown. 
}
\label{sensitivity}
\vspace{+0.5cm}
\end{figurehere}
Note that for a periodic signal at frequency $f$ lasting
for a time interval longer than the observation time $\tau$, we
have simply $n=f\tau$. Then, the signal-to-noise ratio $S/N$ increases 
by the same factor one would obtain comparing $h$ to $h_P$ in equation 
(\ref{eqburstperiodic}). The former approach, i.e. comparing $h_c$
to $h_B$ rather than $h$ to $h_P$, is more general, as it allows 
us to characterize the $S/N$ not only for perfectly periodic signals
($n=f\tau$), or for bursts ($n=1$), but also for events in which the 
emitted frequency shifts to increasingly larger values during 
the spiral-in phase of the binary system. In the latter case,
$n=n(f)$ represents the number of cycles spent in a frequency interval
$\Delta f \simeq f$ around frequency $f$, and hence $h_c$ is the strain 
in a logarithmic frequency interval (Flanagan \& Hughes 1998). 
Typically, the timescale for frequency shift is long compared to 
the wave period, and short compared to the duration of the
observation. Only close to the innermost stable circular orbit (ISCO), the GW 
frequency changes at a rate comparable to the frequency itself ($n\sim 1$ 
and hence $h_c \sim h$). 
In Figure \ref{sensitivity} we also show $h$ and $h_c$ for two 
representative binary systems. 
One should note that the true observable GW signal is, for 
$f> n/\tau$ (the ``knee" frequency in the $h_c$ curves), lower than $h$, 
as for these high frequencies the source is not monochromatic over 
the observation time. 
Such simply consideration naturally leads to define $h_c$ and $h_B$.  

\subsection{Characteristic strain}

Consider now a binary system at comoving distance $r(z)$. The
strain amplitude (sky-and-polarisation averaged) at the rest-frame
frequency $f_r$ is
\begin{equation}
h\,=\,\frac{8\pi^{2/3}}{10^{1/2}}\,\frac{G^{5/3}{\mathcal M}^{5/3}}{c^4r(z)}
\,f_r^{2/3}, \label{eqstrain}
\end{equation}
where ${\mathcal M}=(M_1 M_2)^{3/5}/(M_1+M_2)^{1/5}$
is the ``chirp mass'' of the binary, and all the 
other symbols have their standard meaning. The strain is averaged
over a wave period. The rest-frame energy flux (energy per unit
area per unit time) associated to the GW is
\begin{equation}
\frac{dE}{dAdt}=\frac{\pi}{4}\frac{c^3}{G}f_r^2 h^2.
\label{eqenergyflux}
\end{equation}
As discussed above, the important quantity to consider is the
number of cycles spent in a frequency interval $\Delta f \simeq f$
around a given frequency $f$. Assuming that the backreaction from 
GW emission dominates the orbital decay of a binary, during 
the spiral-in phase one can write
\begin{equation}
n \simeq f_r^2/\dot{f_r}=\frac{5}{96\pi^{8/3}}\,\frac{c^5}{G^{5/3}{\mathcal M}^{5/3}}\,
f_r^{-5/3},
\label{eqenne}
\end{equation}
where we have used the rest-frame frequency shift rate
\begin{equation}
\dot f_r=\frac{df_r}{dt_r}= \frac{96\pi^{8/3}G^{5/3}}{5c^5}{\mathcal M}^{5/3}f_r^{11/3}.
\label{eqdotf}
\end{equation}
Note that $n$ can be computed either in the rest or in the observer frame.
The characteristic strain in an observation of (observed)
duration $\tau$ is then
\begin{equation}
h_c=h\sqrt{n} \simeq \frac{1}{3^{1/2}\pi^{2/3}}\,\frac{G^{5/6}{\mathcal M}^{5/6}}{c^{3/2} r(z)}\,
f_r^{-1/6}, \qquad n<f\tau,
\label{eq1h_c}
\end{equation}
and
\begin{equation}
h_c=h\sqrt{f\tau}\propto f_r^{7/6}, \qquad n>f\tau, 
\label{eq2h_c}
\end{equation}
where $f=f_r/(1+z)$ is the observed frequency.
Using Parseval theorem, it is easy to see that $h_c$ is
related to the Fourier transform of the strain $\tilde h$, as
$h_c^2=2 f_r^2 \tilde h^2(f_r)$, where $\tilde h$ is defined
over the positive frequency axis. The specific energy per unit
area is then
\begin{equation}
\frac{dE}{dAdf_r}=\frac{\pi}{4}\frac{c^3}{G}h_c^2,
\label{eq1specenergy}
\end{equation}
and, from equation (\ref{eq1h_c}), we obtain
\begin{equation}
\frac{dE}{df_r}=\frac{\pi^{2/3}}{3}\,G^{2/3}\,
{\mathcal M}^{5/3}\,f_r^{-1/3}.
\label{eq2specenergy}
\end{equation}
Note that $dE/df_r\propto f_r^{-1/3}$, while (eq. \ref{eqenergyflux}) $dE/dt\propto f_r^{10/3}$.

\subsection{Signal-to-noise ratio}

In an operating interferometer, any stochastic signal will
add up (in quadrature) to $h_B$ to form the effective rms noise of
the instrument, $h_{\rm rms}$. An inspiraling binary is then detected 
if the signal-to-noise ratio
{\it integrated over the observation} is larger than the assumed
threshold for detection, where the integrated $S/N$ is given by
\begin{equation}
S/N_{\Delta f}=\sqrt{ \int_{f}^{f+\Delta f} d\ln f' \, \left[
\frac{h_c(f'_r)}{h_{\rm rms}(f')} \right]^2}. \label{eqSN}
\end{equation}
Here, $f$ is the (observed) frequency emitted at the starting time $t=0$ 
of the observation, and $\Delta f$ is the (observed) frequency
shift in a time $\tau$ starting from $f$. The latter is implicitly
given by
\begin{equation}
\tau=\int_{f}^{f+\Delta f} \frac{df'}{\dot f'}. \label{eqfmax}
\end{equation}
where $df/\dot f=(1+z)df_r/\dot f_r$. 
The frequency at the ISCO is, strictly speaking, defined only in the test 
particle limit $M_2 \ll M_1$. In the general case, various estimate of the transition point from 
in-spiral to plunge exist, and differ by a factor of 3 at most (e.g., Kidder, Will \& Wiseman 1993; Cook 1994). 
Such uncertainties do not affect our results in any manner, so we use, for the observed frequency at the ISCO, 
the conventional Keplerian defintion:
\begin{equation}
f_{\rm ISCO}=\frac{c^3}{6^{3/2}\pi G}\frac{1}{(M_1+M_2)}\,(1+z)^{-1}.
\label{eqfisco}
\end{equation}
Replacing $f+\Delta f$ with $f_{\rm ISCO}$ in equation (\ref{eqfmax})
gives $\tau_{\rm ISCO}$, the time needed to span the frequency
interval $[f,f_{\rm ISCO}]$, to be compared {\it a priori}
to $\tau$. In the case $\tau>\tau_{\rm ISCO}$, we then set
$f+\Delta f=f_{\rm ISCO}$ in equation (\ref{eqSN}).
In Figure \ref{singlesource} we plot $h_c$ for different MBHBs 
at different redshifts, compared to the {\it LISA} $h_{\rm rms}$ 
(see \S~4.1) multiplied by a factor of 5, assuming a 3-year observation. 
If $h_c > 5h_{\rm rms}$, then the signal has, approximately, an 
integrated $S/N>5$. This is for illustrative purposes only, as the 
actual $S/N$ must be integrated 
over the observing period using equation (\ref{eqSN}). 

At frequencies higher than the ``knee'', the time spent around a given
frequency is less than 3 years, and $h_c\propto f^{-1/6}$. The
signal shifts toward higher frequency during the
observation, and reaches the ISCO and the coalescence
phase in most cases. The lowest curve represents a low mass, high redshift equal
mass binary. As we shall see below, these sources are common in our
hierarchical model for MBH assembly. In terms of their detectability
by {\it LISA}, 
they represent a somewhat different class of events. Contrary to
the case of more massive binaries present at lower $z$, the final
coalescence phase of light binaries lies at too high frequecies, 
well below the {\it LISA} threshold. 

For frequencies much below the knee, the characteristic strain is 
proportional to $f^{7/6}$, as the timescale for frequency shift is 
longer than 3 years.
The signal amplitude is then limited by the observation time, not
by the intrinsic properties of the source. The source will be
observed as a ``stationary source", a quasi-monochromatic wave for
the whole duration of the observation. An increase in the observation
time will result in a shift of the knee toward lower frequencies. The
time needed for the sources to reach the ISCO starting from the
knee frequency is, approximatively, the observing time.
Figure \ref{singlesource} shows that very few stationary sources 
above threshold should be expected anyway.
\begin{figurehere}
\vspace{0.5cm}
\centerline{\psfig{figure=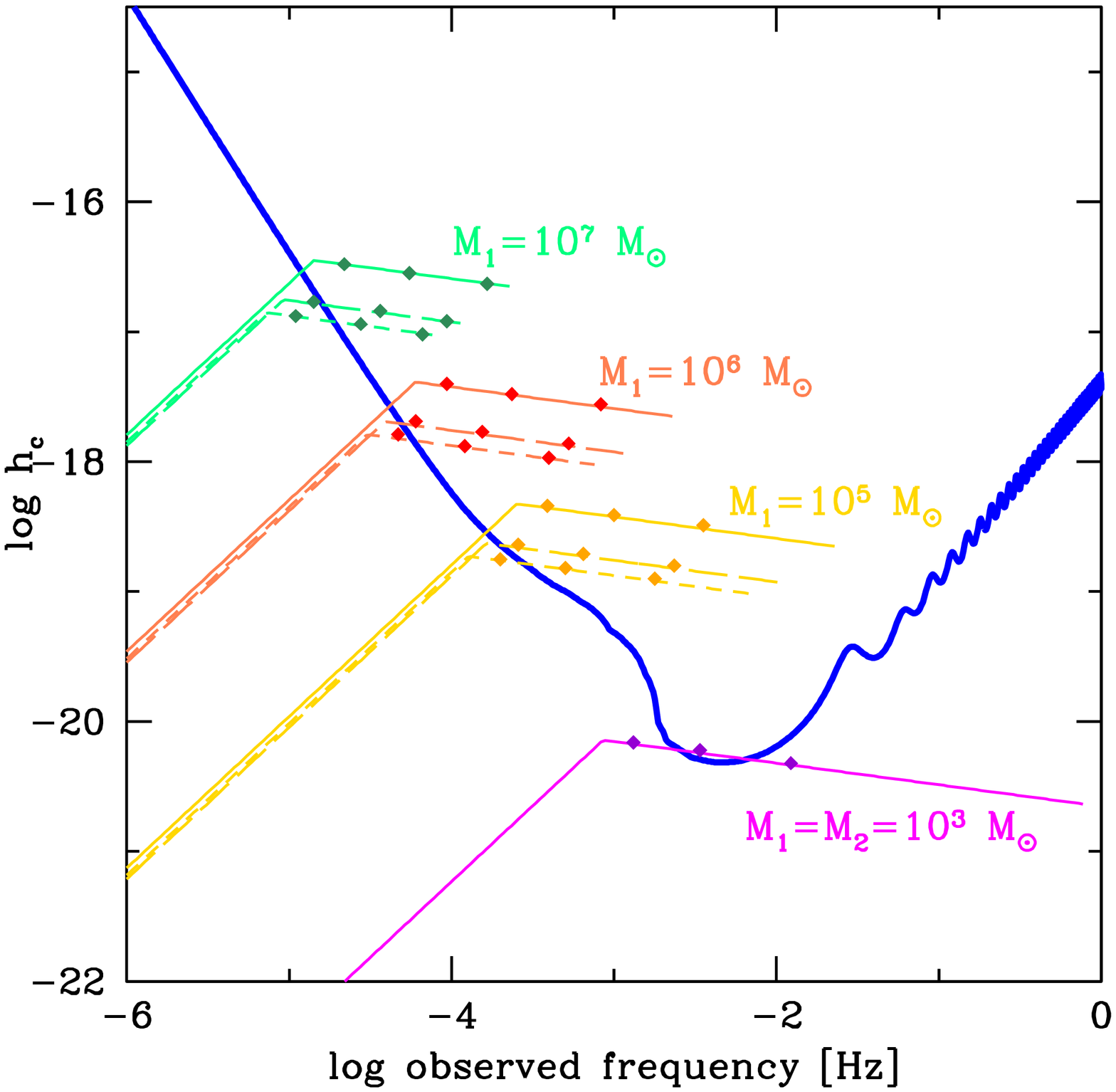,width=3.6in}}
\vspace{-0.0cm}
\caption{\footnotesize 
Characteristic strain $h_c$ for
MBHBs with different masses and redshifts. From
top to bottom, the first three curves refer to systems with
$\log(M_1/M_{\odot})=7,6,5$, respectively, and
$M_2=0.1 M_1$. The solid, long-dashed, and short-dashed lines
assumes the binary at $z=1,3,5$, respectively. A 3-year
observation is considered. The lowest solid curve assumes an equal mass
binary $M_1=M_2=10^3 M_{\odot}$ at $z=7$. The small diamonds on
each curve mark, from left to right, the observed frequency at 1
year, 1 month
and 1 day before coalescence. The thick curve is {\it LISA} $5h_{\rm
rms}$ (see \S~4.1), approximatively the threshold for detection
with $S/N \geq 5$.
}
\label{singlesource}
\vspace{+0.5cm}
\end{figurehere}

\subsection{Coalescence rate and number counts}

Given a coalescence rate $R$, using the frequency shift rate
$\dot f$ (eq. \ref{eqdotf}), we can solve for the mean number of
{\it individual} binaries resolved during an observing period $\tau$.
We begin considering that a MBHB spans, during its lifetime, a finite 
frequency range, $f_{\rm min}<f<f_{\rm ISCO}$, where
the lower limit is set to the observed frequency at the hardening
radius (Quinlan 1996). Then, from continuity, the number of individual 
observable MBHBs can be computed as
\begin{equation}
N_{\tau}=R\int_{f_{\rm min}}^{f_{\rm ISCO}}\frac{df}{\dot
f}\,\,+\,\,R \tau.
\label{eqbinsky}
\end{equation}
The first term is simply the integrated density of sources in the
frequency domain, and does not depend on $\tau$. It is the number
of sources caught in a snapshot of the entire sky. The second term
is the number of new binaries born (at frequency $f_{\rm min}$) during the
observation time $\tau$, and must be equal to the number of
coalescences within the same period.

The general argument above does not consider that real detections must be above a 
specified minimum $S/N$, where the $S/N$ is given by
equation (\ref{eqSN}). Including a threshold criterium, the number of MBHBs 
with $S/N>s$ in an observation of duration $\tau$ is then
\begin{equation}
N_{\tau}(>s)=R\int_{f_{\rm min}}^{f_{\rm ISCO}}\frac{df}{\dot
f}H_s(\Delta f)\,\,+\,\,R \int_{f_{\rm min}}^{f_{\rm max}}\frac{df} {\dot
f}H_s(\Delta {f_{\rm min}}), \label{eqbinSN}
\end{equation}
where
\begin{equation}
H_s(\Delta f)=\cases{1,\,\,\,\, S/N_{\Delta f}\geq s \cr 0,\,\,\,
 S/N_{\Delta f}< s}.
\end{equation}
In the second term of equation (\ref{eqbinSN}), which again accounts for
the new binaries formed at the hardening radius, $f_{\rm max}$ is the
frequency reached after 3 years starting from $f_{\rm min}$, and the function
$H_s(\Delta {f_{\rm min}})$ is evaluated by integration of the $S/N$ from
$f_{\rm min}$ to $f$. Given the exceedingly low value of $f_{\rm min}$, this
second term is totally negligible for an experiment such as {\it LISA}.

\section{Hierarchical growth of massive black holes}

The theory and method outlined in the previous sections allow us
to fully characterize the expected contribution of MBHBs 
in the spiral-in phase to the {\it LISA} data stream, once the coalescence 
rate of MBHBs is specified. In this work a hierarchical structure 
formation scenario for the assembly and growth of MBHs in which seed holes 
form far up in the dark halo ``merger tree'' is assumed. We use exactly 
the same model discussed in Volonteri \etal (2003a, 2003b) and in Paper I.
Its main features are briefly summarized in this section.  

We track backwards the merger history of 220 parent
halos with present-day masses in the range $10^{11}-10^{15}\,\msun$ 
with a Monte Carlo algorithm based on the extended Press-Schechter formalism
(see, e.g., Cole \etal 2000). 
Seed holes with $m_{\rm seed}=150\msun$ are placed within rare 
high-density regions (minihalos) above the cosmological Jeans and cooling 
mass at redshift 20. Their evolution and growth is followed through Monte 
Carlo realizations of the halo merger hierarchy combined with semi-analytical
descriptions of the main dynamical processes, such as dynamical friction
against the dark matter background, the shrinking of MBHBs via 
three-body interactions, their coalescence from the emission of 
gravitational waves, triple MBH interactions, and the effect of gravitational 
recoil. 
\begin{figurehere}
\vspace{0.5cm}
\centerline{\psfig{figure=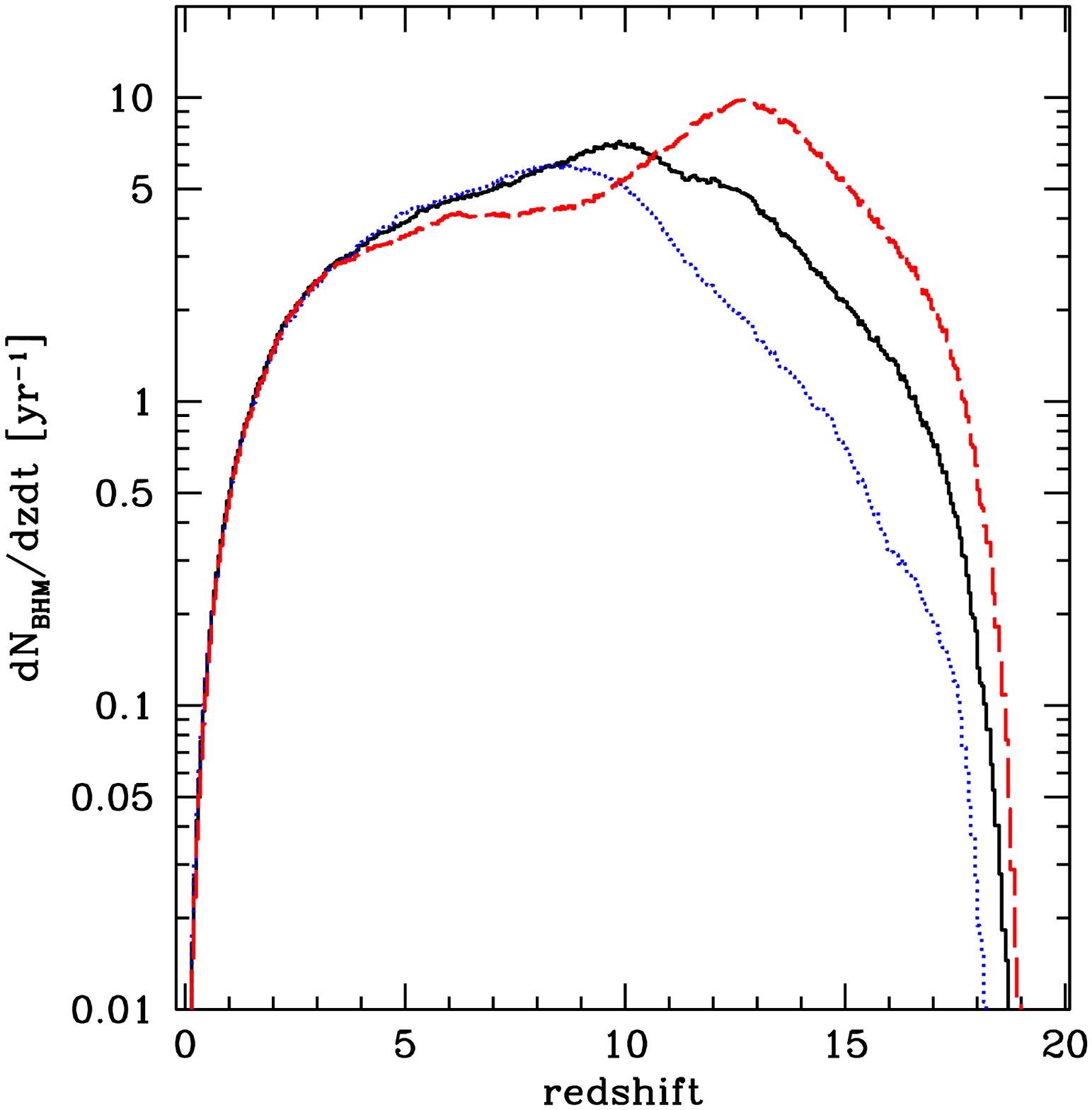,width=3.6in}}
\vspace{-0.0cm}
\caption{\footnotesize
Number of coalescences of MBHBs observed per year at $z=0$ per unit redshift.
Our fiducial rate ({\it thick solid line}) is compared to a case in which the 
hardening timescale is increased by a factor of 3 ({\it dotted line}) 
or reduced by the same factor ({\it dashed line}).
}
\label{figrate}
\vspace{+0.5cm}
\end{figurehere}
Quasar activity is triggered during major mergers. 
We assume that the more massive hole accretes, at the Eddington rate, a gas 
mass fraction that scales with the fifth power of the host halo circular 
velocity (Ferrarese 2002). 

In a typical merger event, dynamical friction drives the satellite halo 
toward the centre of the new forming system, leading to the formation
of a bound MBHB in the violently relaxed stellar core.
As the binary separation decays, the effectiveness of 
dynamical friction slowly declines; the bound pair then hardens by capturing 
stars passing  within a distance of the order of the binary semi-major 
axis and ejecting them at much higher velocities (gravitational slingshot).
The  heating of the surrounding stars by a decaying MBH pair creates 
a low-density core out of a preexisting stellar cusp, slowing down further 
binary hardening (see, e.g., Milosavljevic \& Merritt 2001). If the 
hardening continues sufficiently far, GW emission takes over, driving
the pair to coalescence. Figure \ref{figrate} shows the number of 
MBHB coalescences
per unit redshift per unit observed year predicted by our model: we expect
$\sim 60$ coalescences per year, the vast majority involving quite light
binaries ($M_1+M_2\leq10^5\msun$).
The model was shown to reproduce rather well the observed luminosity function 
of optically-selected quasars in 
the redshift range $1<z<5$ and the evolution of the nuclear MBH mass 
density with cosmic time (Volonteri \etal 2003a), and to provide a 
quantitative explanation 
to the stellar cores observed today in bright ellipticals as a result of the 
cumulative eroding action of shrinking MBHBs (Volonteri \etal 2003b).
\begin{figurehere}
\vspace{0.5cm}
\centerline{\psfig{figure=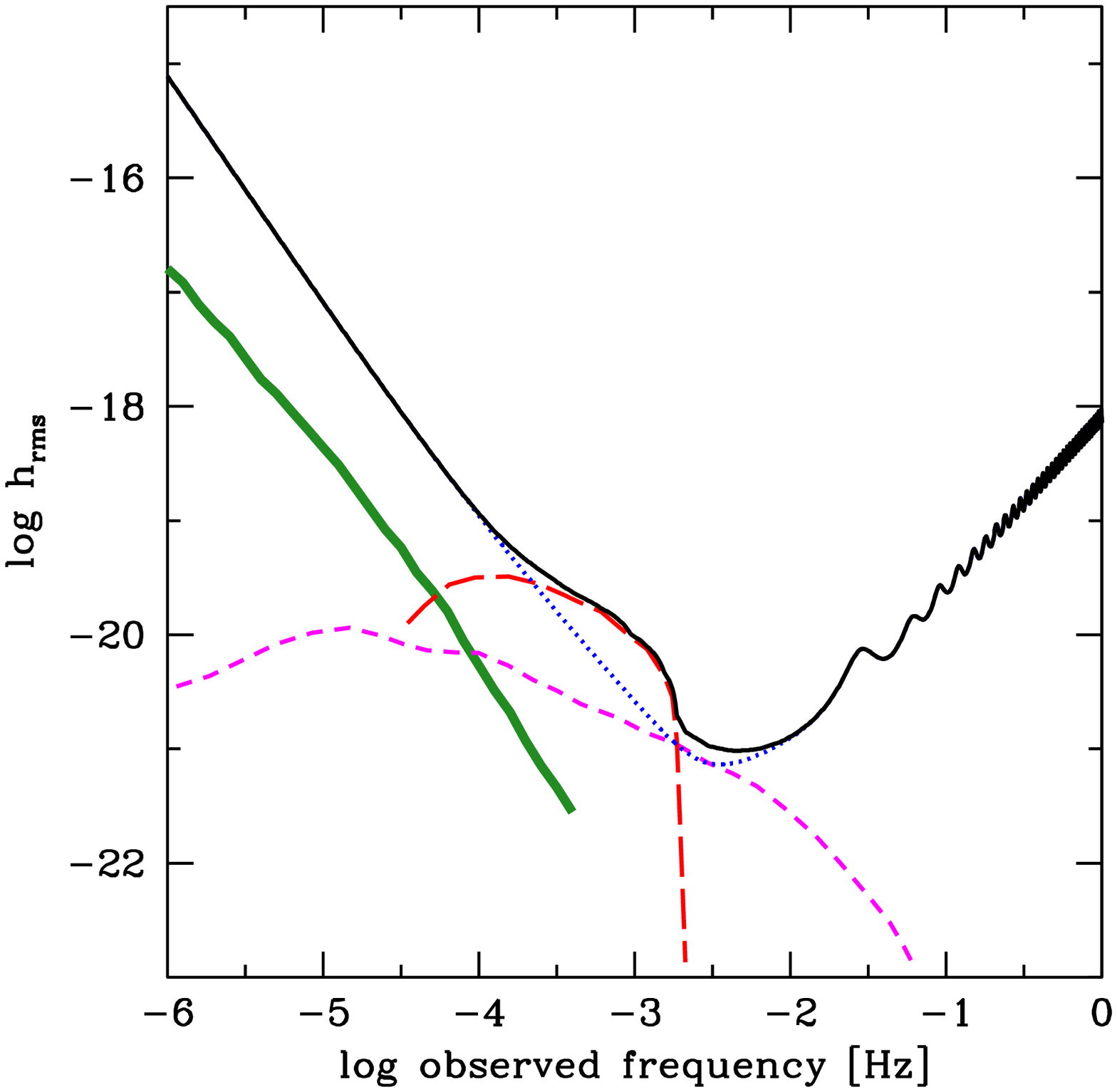,width=3.6in}}
\vspace{-0.0cm}
\caption{\footnotesize
Estimated {\it LISA} rms confusion noise ({\it solid line}), as
the quadratic sum of the {\it LISA} instrumental single-arm Michelson
noise $h_B$ ({\it dotted line}), the confusion noise from
unresolved galactic (Nelemans \etal 2001, {\it long-dashed line}), and
extragalactic (Farmer \& Phinney 2003, {\it short-dashed line})
WD-WD binaries, and our estimate of the confusion noise
from unresolved MBHBs ({\it thick-solid line}).
}
\label{noise}
\vspace{+0.5cm}
\end{figurehere}

\section{Number counts}

\subsection{Stochastic noise from MBHBs}

The customary definition of GW confusion noise level is the amplitude
at which there is, on average, at least one source per frequency
resolution bin. The frequency bin width is $\Delta f=1/\tau$, so
the longer the observation, the smaller the noise. As pointed out
by Cornish (2003), the crude ``one bin rule" is much too simple to
properly describe a binary system. Using detailed information theory, 
Cornish (2003) shows that a GW background
becomes unresolvable when there is, on average, at least one
source per eight bins.

In the last decade a considerable effort has gone into
quantifying the galactic and extragalactic confusion noise in
the band 0.01 mHz - 1 Hz (e.g., Schneider \etal 2000; Freitag 2001; 
Nelemans \etal 2001; Farmer \& Phinney 2003).
We have then applied the ``eight bin rule" to asses the confusion
noise associated with the evolving population of MBHBs, compared to the
most recent estimates of the noises from galactic (Nelemeans \etal
2001, ``one bin" rule, 1 year observation) and extragalactic (Farmer \& Phinney 2003, 
``one-bin" rule, 3-year observation) white dwarf (WD) binaries. 
As shown in Figure \ref{noise}, MBHBs produce confusion noise at $f\lsim 4\times 10^{-4}$ Hz. 

Figure \ref{noise} also shows the global {\it LISA} $h_{\rm rms}$, 
along with separate contributions from different source populations. 
Though, as expected, MBHB stochastic noise dominates
over WD-WD signals at low frequencies, it lies more than an order of
magnitude below the instrumental {\it LISA} sensitivity curve, and hence
its contribution to the {\it LISA} $h_{\rm rms}$ can be ignored. 
On the other hand, this hampers the possibility that {\it LISA} could take 
advantage of the MBHB noise to probe the 
cosmological evolution of such particular parent population.

\subsection{Mass function and redshift distribution}

We have divided the resolved sources into ``merging"
and ``in-spiral" binaries (MBs and IBs, respectively). 
The former are those binaries that reach the ISCO during the duration 
of the observation with a signal above threshold. 
\begin{figurehere}
\vspace{0.5cm}
\centerline{\psfig{figure=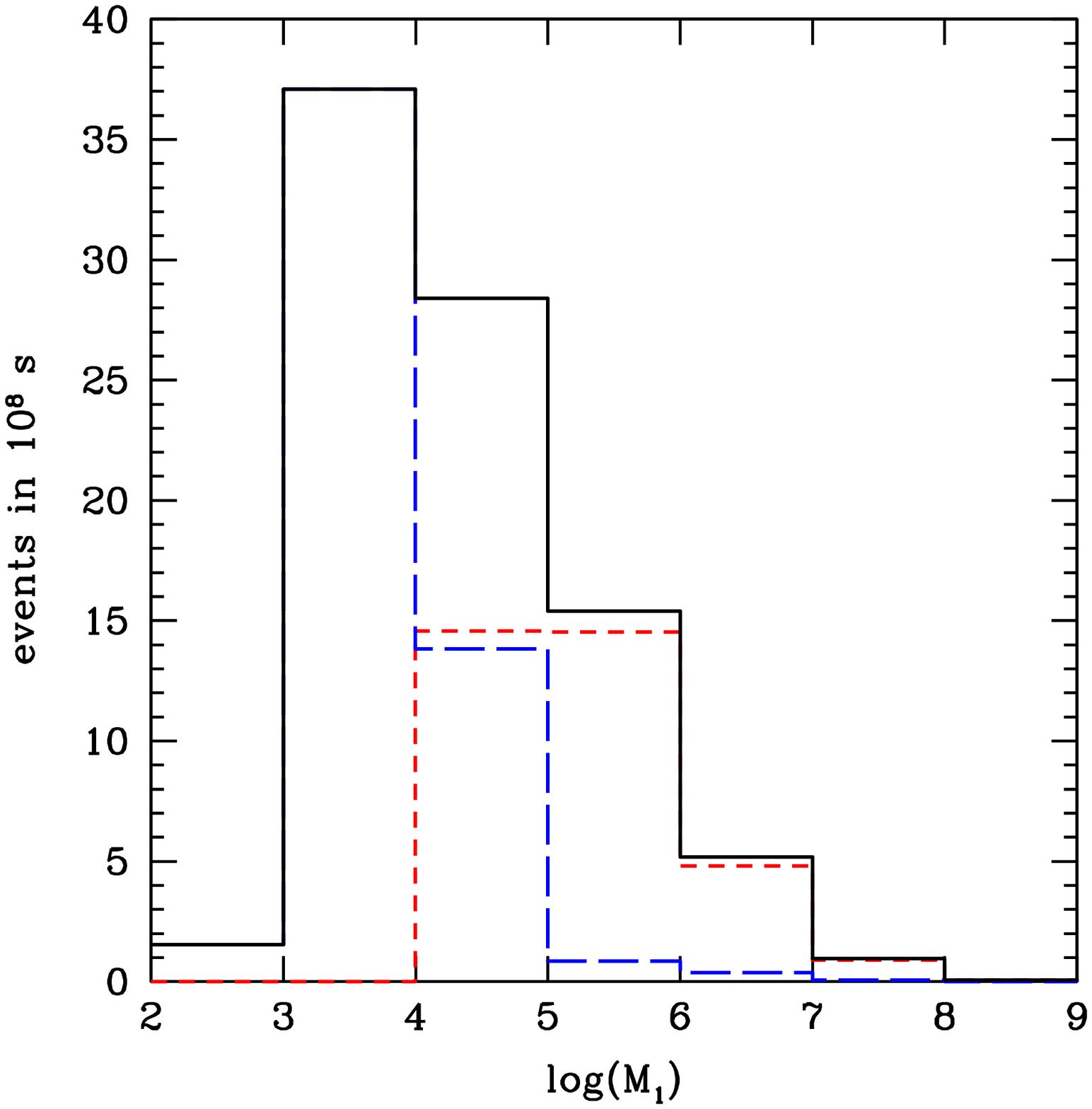,width=3.6in}}
\vspace{-0.0cm}
\caption{\footnotesize
Mass distribution of the more massive member of MBHBs resolved with $S/N>5$ by 
{\it LISA} in a 3-year mission ({\it solid line}).
The separate counts for MBs ({\it short-dashed line}) and IBs 
({\it long-dashed line}) are also shown. 
}
\label{massfunction}
\vspace{+0.5cm}
\end{figurehere}
These events are of particular importance, as they probe strong field
effects and represent a unique chance of observing the
coalescence and ring-down phases of MBHBs. Resolved IBs, 
instead, do not allow a direct observation of the
coalescence phase. These events arise from light binaries
whose final coalescence phase lies below threshold, and from  
binaries of all masses with $\tau_{\rm
ISCO}>3$ years at the very start of the observation. We expect very few
IBs in this last stage anyway, because binaries above
threshold have typically $\tau_{\rm ISCO}<3$ years when the
observation starts (see Fig. \ref{singlesource}). An example of
a resolved IB is represented by the lowest curve of
Figure \ref{singlesource}. MBHBs in this class have an
integrated $S/N$ above threshold, though the coalescence phase
occur at too high frequency to be directly observed by {\it LISA}.
\begin{figurehere}
\vspace{0.5cm}
\centerline{\psfig{figure=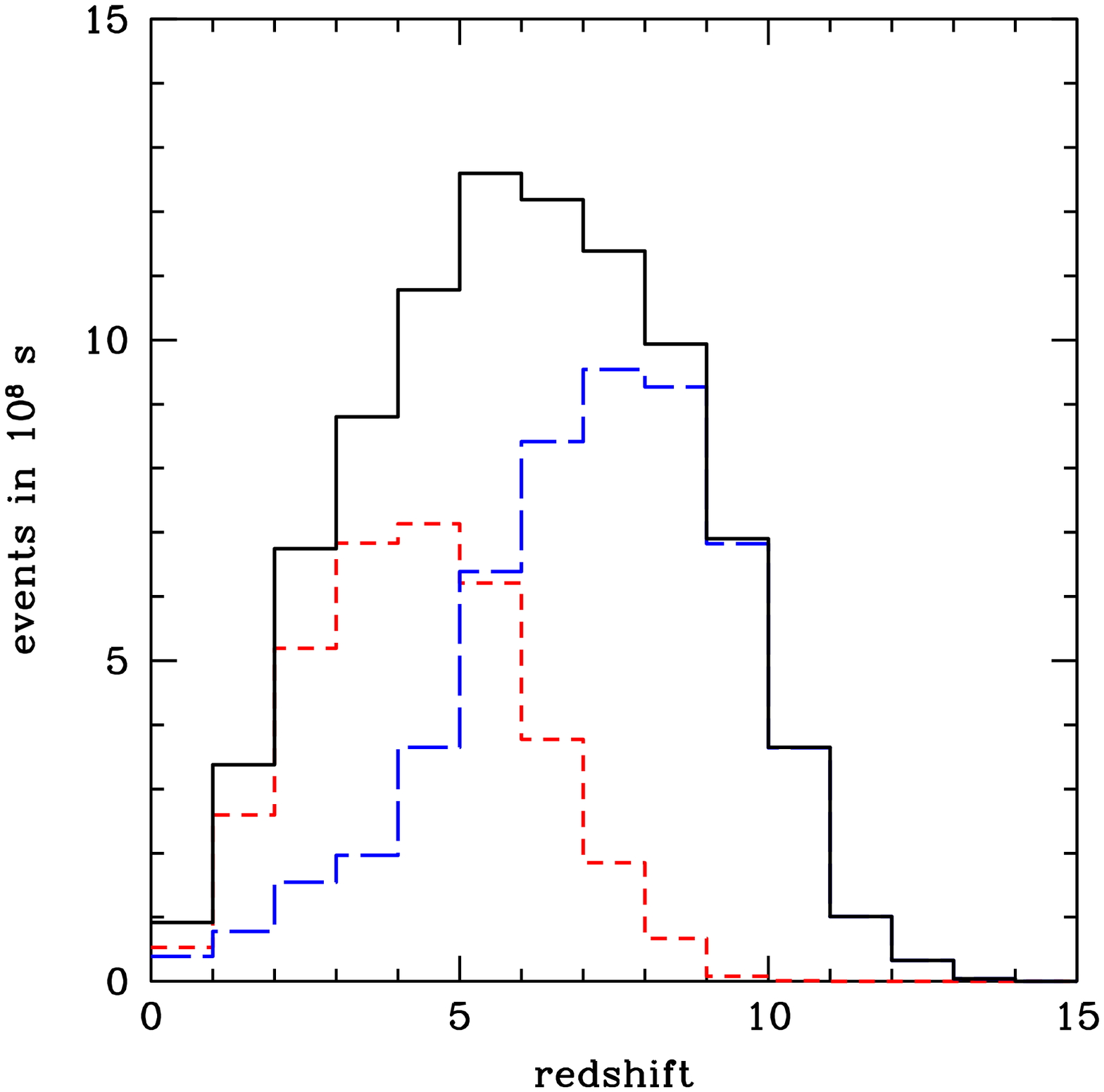,width=3.6in}}
\vspace{-0.0cm}
\caption{\footnotesize
Differential redshift distribution of MBHBs resolved with $S/N>5$ by {\it LISA} in a 
3-year mission. Line style as in Fig. \ref{massfunction}.
}
\label{diffcounts}
\vspace{+0.5cm}
\end{figurehere}
An obvious consequence of our classification is that MBs 
have larger mass and a lower redshift than IBs. The mass
distribution of the most massive member of the binary $M_1$ 
is shown in Figure \ref{massfunction}. The differential and
cumulative redshift distributions are plotted in Figure
\ref{diffcounts} and Figure \ref{cumcounts}, respectively.
Detectable IBs are $\sim 55$ in total, and account for almost
all of the MBHBs observed at $z\gsim 7$. Conversely, the rarer
MBs ($\simeq 33$ in total) are confined to the 
redshift interval $2\lsim z \lsim 7$.
Figure \ref{massratio} shows the average mass ratio
$M_2/M_1$ for MBs and IBs. As expected, this is larger for IBs, given their
larger average redshift. The ratio decreases at low
redshift, as a consequence of the complicated merger history of host
dark matter halos. The reason why the average mass ratio of MBs 
peaks at $M_2/M_1 \simeq 0.15$ lies in the fact 
that both the probability of halo mergers (because of the steep P\&S halo mass 
function) 
and the dynamical friction timescale increase with
decreasing halo mass ratio. Hence, fast equal mass mergers are rare, while 
in more common unequal mass mergers it takes longer than an Hubble time to 
drag the satellite hole to the centre.   
\begin{figurehere}
\vspace{0.5cm}
\centerline{\psfig{figure=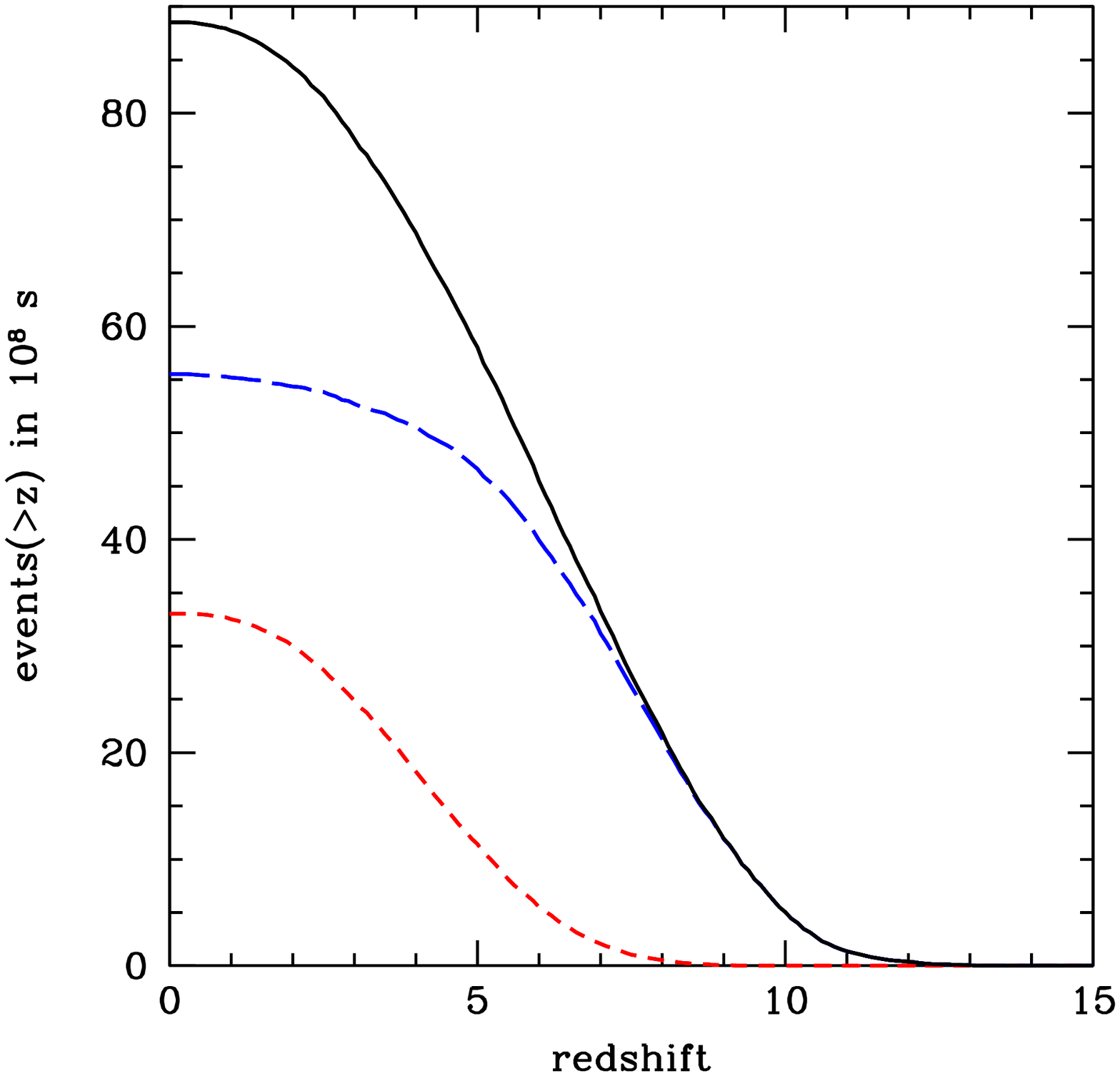,width=3.6in}}
\vspace{-0.0cm}
\caption{\footnotesize
Cumulative (integral from z to $\infty$)
redshift distribution of MBHBs resolved with $S/N>5$ by {\it LISA} in a 3-year mission.
Line style as in Fig. \ref{massfunction}.
}
\label{cumcounts}
\vspace{+0.5cm}
\end{figurehere}

\begin{figurehere}
\vspace{0.5cm}
\centerline{\psfig{figure=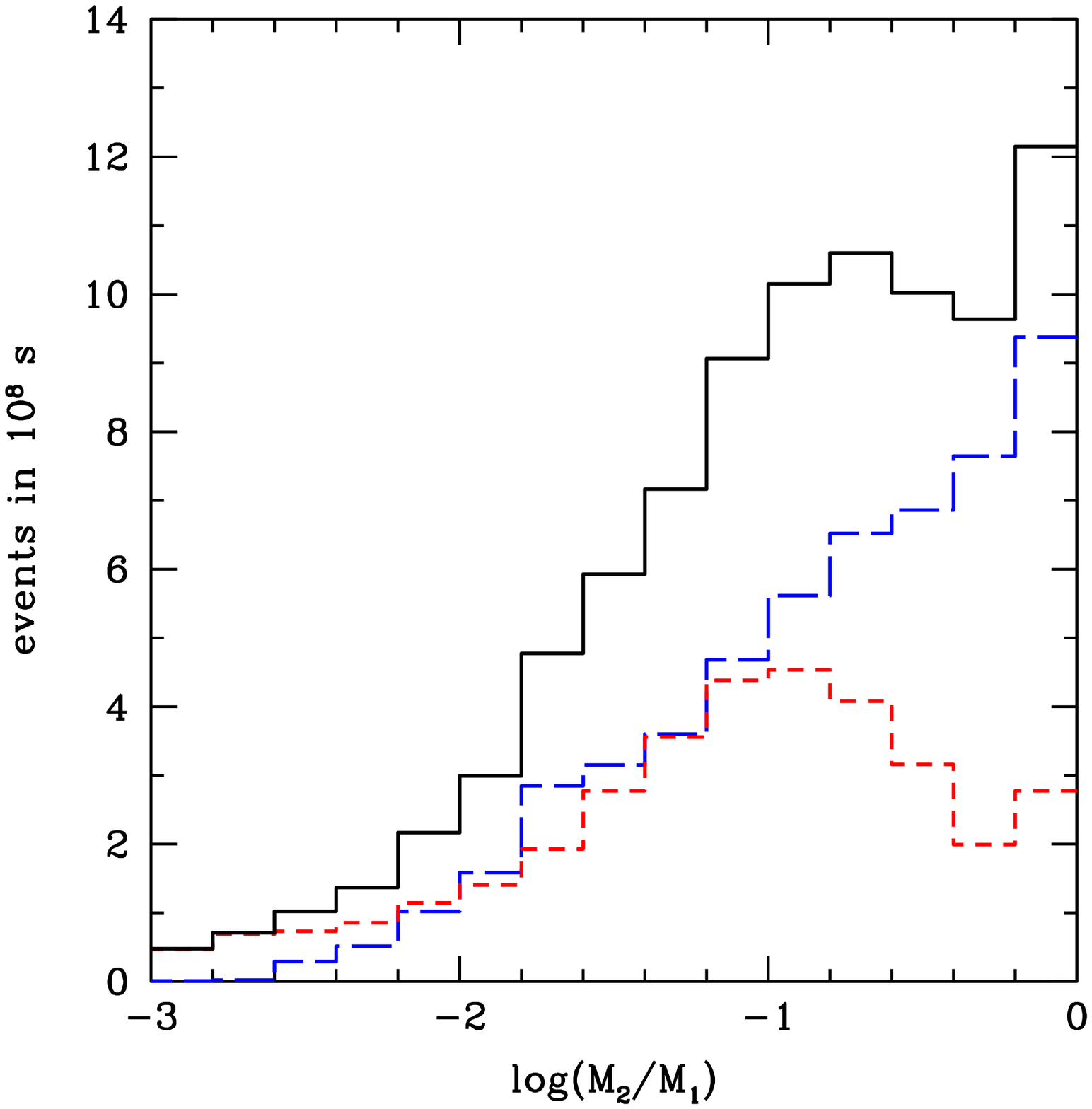,width=3.6in}}
\vspace{-0.0cm}
\caption{\footnotesize
Mass ratio distribution of MBHBs resolved with $S/N>5$ by {\it LISA} in a 3-year mission.
Line style as in Fig. \ref{massfunction}.
}
\label{massratio}
\vspace{+0.5cm}
\end{figurehere}
We can define the detection efficiency of a specific mission 
as the number of observable events divided by the expected number 
of coalescences in the same
time interval. Figure \ref{eff} shows the global (MBs+IBs) detection 
efficiency for {\it LISA} and the efficiency considering as ``detections" only MBs.
The large GW-brightness of MBHBs is such that {\it LISA} will 
observe $\gsim 90$\% of all
coalescences occurring at $z\lsim 5$. The efficiency falls below $0.5$ 
only for MBHBs at $z\gsim 8$. The efficiency to MBs only is, obviously, lower. 
Figure \ref{eff} shows that a space--based interferometer such as {\it LISA}
can directly observe the final stage of the spiral-in phase
of about half of all MBHBs coalescing at $z\simeq 5$.
\begin{figurehere}
\vspace{0.5cm}
\centerline{\psfig{figure=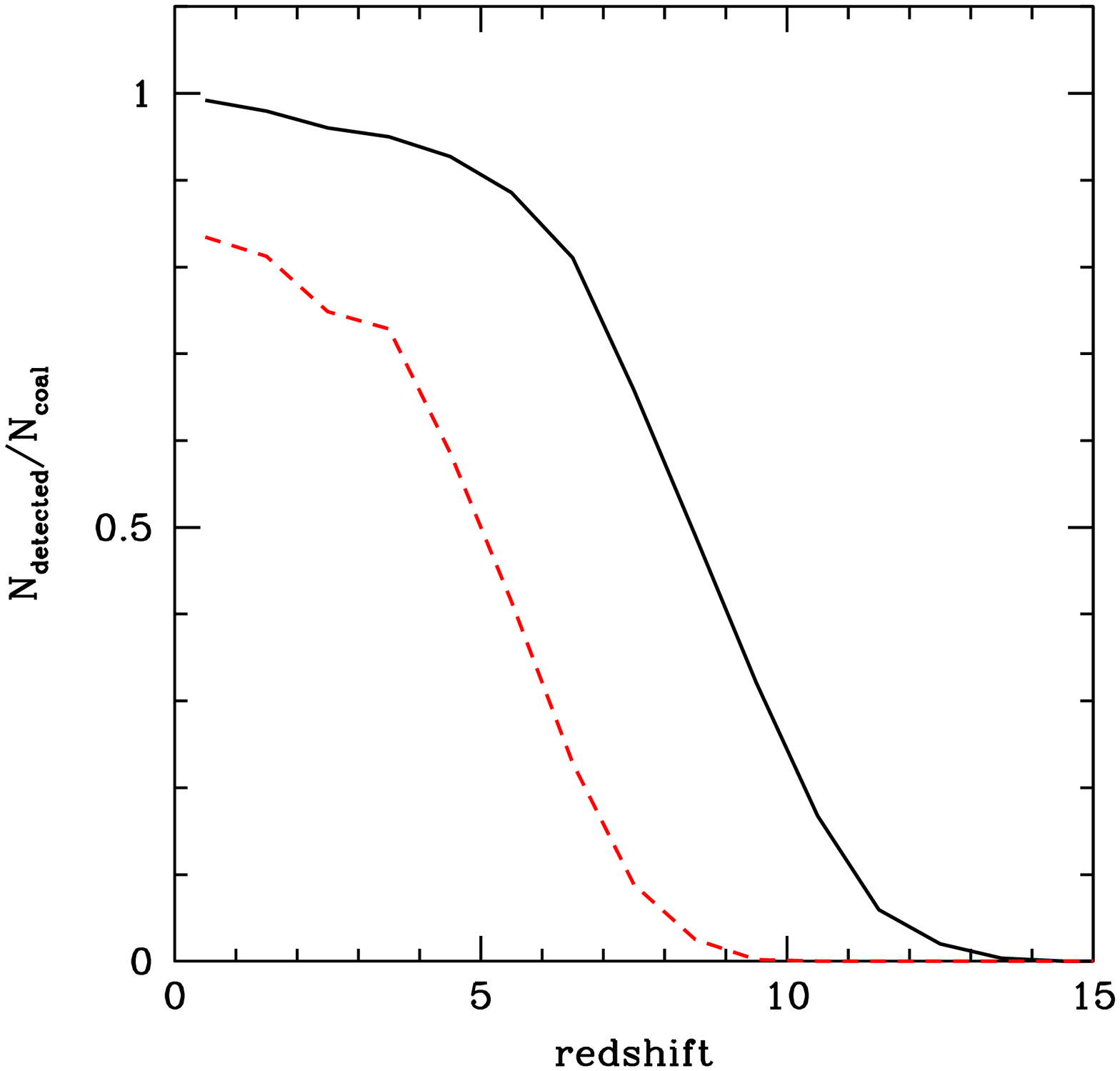,width=3.6in}}
\vspace{-0.0cm}
\caption{\footnotesize
Detection efficiency ({\it solid line}),
defined as the number of detected events (MBs+IBs) divided by the total
number of coalescences in the same time interval, as a function of redshift.
The efficiency considering only MBs as detections is also shown ({\it dashed line}).
}
\label{eff}
\vspace{+0.5cm}
\end{figurehere}

\section{Discussion}

In this paper we have characterized the GW signal produced by cosmological 
MBHBs, and we have then folded the signal into the {\it LISA} performance 
capabilities. We find that {\it LISA} should resolve 
more than 90\% of all cosmological coalescences of MBHBs occurring at 
$z\lsim 5$. The detection efficiency is already $\gsim 0.5$ for MBHBs at 
$z\simeq 8$. We showed that the confusion noise from residual unresolved 
MBHBs is expected to be at least an order 
of magnitude below {\it LISA} instrumental noise.

We have divided the resolved events into merging and in-spiral binaries.
MBs are associated with systems at relatively low redshift involving
heavy pairs ($10^4-10^7$ M$_{\odot}$). Their strong GW signal can be used  
to study the orbital evolution of the pair until the ISCO, allowing to test
GR in extreme conditions. On the contrary, IBs are less massive pairs at 
higher redshift. Such systems can be generally observed, with moderate 
integrated $S/N$, only for a relatively short amount of time, from few weeks 
to few months, before the ISCO is reached. IBs are nevertheless important 
as they push the limit of observable MBHBs out to $z\simeq12$, and allow 
studies of the formation and assembly of seed holes of intermediate masses.
Binaries at even earlier epochs, while common in our model, are 
unfortunately too light to be observed by {\it LISA} with a relevant $S/N$. 
 
In our refined study, the $\sim 20$  ``stationary sources'' discussed 
in Paper I, i.e. sources whose shift in frequencies $\Delta f$ during 
the observing period is $\lsim f$, appear to be undetectable. 
This is because a too optimistic
sensitivity threshold for {\it LISA} was assumed in Paper I for frequencies 
below 0.1 mHz, where 
most of the stationary sources are expected. Indeed, an order of 
magnitude improvement of the 
{\it LISA} sensitivity below 0.1 mHz would lead to the detection of
$\sim 100$ of such events. We stress here that our results may be sensitive 
to different model 
parameters and assumptions. As done in Paper I, we explore such possibility 
by running test models in which the hardening timescale $t_h$ was divided
and multiplyied 
by a factor of 3. The resulting coalescence rates are plotted in Figure 
\ref{figrate}. We find 50 coalescences per observed year
in the ``slow hardening case" and 78 in the ``fast hardening" case, 
compared to 64 in our reference model. 
In terms of {\it LISA} number counts, the effects are small. 
In the slow hardening case, the coalescence rate decreases at 
$z\gsim 9$ as a large fraction of binaries has $t_h$ longer than the 
then Hubble time $t_H$. At lower redshifts it is $t_h<t_H$ anyway, 
the coalescence rate is basically unaffected with respect to the 
fiducial case, and so are the number counts (both for MBs and IBs). 
To obtain a significative reduction of observable sources, $t_h$ must 
increase by a larger factor, so that also MBHBs in the range $5\lsim z 
\lsim 10$ will have $t_h\gsim t_H$. By increasing $t_h$ by an 
order of magnitude, we find the number of coalescences per observed 
year decreasing to 30: in a 3-year observation {\it LISA} would detect 
$20$ ($40$) MBs (IBs). 
Note that, as increasing $t_h$ results in a lower average redshift 
of coalescences, this also increases the global detection efficiency. 
In other words, there are less sources, but a larger fraction of them is 
detectable.  

In the fast hardening case, more binaries can coalesce at early times, 
and the number of surviving MBHBs at $6\lsim z\lsim 12$ decreases. 
This ultimately causes a slight reduction in the number of IBs observable by 
{\it LISA} (49 in 3 years, compared to 55). 
From Figure \ref{figrate}, it is also evident that, for $z\lsim 5$, the 
coalescence rate 
is almost identical to the standard case, implying that the number of 
detectable MBs will remain approximately the same.

In Figure \ref{massratio}, we showed that IBs have an average  
mass ratio higher than MBs. To check whether this result depends on  
our assumption of equal-mass seed holes, we ran a test model with a flat 
initial mass function for the seed, $10<m_{\rm seed}<500$M$_{\odot}$.
The resulting binary mass distribution relevant for {\it LISA} was 
basically unaffected.

The vast majority of IBs are low mass systems at fairly high 
redshift. Their characteristic strain lies just above the {\it LISA} 
threshold at frequencies of the order of $10^{-3}-10^{-2}$ Hz (see Fig. 
\ref{singlesource}) where confusion noises from unresolved galactic 
and extragalactic WD-WD binaries dominate
the sensitivity curve (see Fig. \ref{noise}). WD-WD confusion 
noise levels are difficult to compute because of the many 
uncertainties in stellar population synthesis models, and in 
estimating the fraction of binary stars in galaxies. 
In our fiducial sensitivity curve we have added to the {\it LISA} 
effective noise the galactic WD-WD confusion noise computed by Nelemans 
et al. (2001), and the extragalactic WD-WD confusion noise estimated
by Farmer $\&$ Phinney (2003). Note that Nelemans \etal (2001) assumed 
1, rather than 3, year integration, and both estimates employ 
the one bin rule. Using the eight bin rule we expect these noises to 
increase to some extent, but differences should be small.
An alternative accurate estimation for galactic WD-WD confusion noise was 
performed by Hils $\&$ Bender (2000), who assumed a three bin rule. Using their
noise level, which is somewhat higher than that computed by Nelemans 
et al. (2001), the number of IBs observed by {\it LISA} in 3 years slightly decreases, from 55
to 50. 

Compared to unresolved galactic WD-WD binaries, the uncertainties in the extragalactic
WD-WD confusion noise are much larger, and may have a more relevant impact on number counts.  
The estimate of Schneider et al. (2000) lies nearly a factor of four 
above Farmer \& Phinney's ``fiducial'' model, and this leads to a significant
reduction of observable IBs, from 55 to 44.
The lowest estimate we could find in the literature was Farmer \& Phinney's
``pessimistic'' model, which is about a factor of four lower than 
their fiducial one; in this case we count 63 observable IBs. 

Another potential source of noise in the frequency range 1-10 mHz is            
captures of compact objects (white dwarfs, neutron stars, and stellar-mass
black holes) by MBHs in galaxy centers. Capture rates are quite uncertain, 
and estimates of relevant confusion noises span more than an order of 
magnitude in $h_c$ (see Barack \& Cutler 2004 for a detailed discussion). 
We have estimated the impact of compact object captures on {\it LISA} 
number counts assuming the more optimistic rates calculated by       
Freitag (2001). The number of observable IBs decreases in this case to 43. 
More conservative rate estimates do not affect appreciably the counts. 

To summarise, in the context of our model, 
we can assign an approximate error of $\simeq 50$\% to the 
number of high-$z$ MBHBs detectable by {\it LISA}. We conclude remarking 
that the bulk of detections involves binaries with masses in the 
interval $10^3-10^5 \msun$, a range where black holes have never been 
observed. Genuine supermassive BH binaries, whose existence is more secure 
on observational grounds, appear too ``heavy'' for interferometers working 
in the band 0.01 mHz - 1 Hz.

\acknowledgments 
\noindent
We thank P. Bender and A. Vecchio for several enlightening
discussions during the preparation of this paper.
Support for this work was provided by NASA grant NNG04GK85G and 
by NSF grant AST-0205738 (P.M.).

{}

\end{document}